\documentclass[runningheads]{llncs}
\usepackage{graphicx}
\usepackage{amsmath}
\usepackage{amsfonts}

\usepackage{tikz-cd} 
\usepackage{hyperref}
\begin{document}
\title{Nash embedding: a road map to realizing quantum hardware}
%
%
\author{Faisal Shah Khan}
%
\authorrunning{Faisal Shah Khan}
%
\institute{Quantum Computing, Inc., Leesburg, VA 20175 USA, \\
Department of Mathematics, Khalifa University, Abu Dhabi UAE\\
Email: fkhan@quantumcomputinginc.com }
\maketitle              

\begin{center}
\today
\end{center} 

\begin{abstract}
The non-Euclidean nature of the mathematical model of quantum circuits leaves open the question of their practical implementation in hardware platforms which necessarily reside in the Euclidean space $\mathbb{R}^3$. On the other hand, reversible circuits are elements of Euclidean spaces, making their physical realization in hardware platforms possible and practical. Here, the quantum circuit model for quantum computing is mapped into that of reversible computing in a mathematically robust fashion using Nash embedding so that every quantum computation can be realized as an equivalent reversible one.
\end{abstract}

\section{Introduction}

Now that the industry has started producing and making available software solutions that will run on (or will be compatible with) hardware implementing quantum computations, the time is right to start asking whether mathematically formal and robust road-maps to engineering these hardware exist. While I will answer this question in the affirmative in this letter by proposing Nash embedding as one such road-map, I will also raise questions relating to potential applications of Nash embedding that intuitively appear to have a positive answer.

Prototype quantum computing systems are currently available from several vendors, including D-Wave Systems and Rigetti Computing (and IBM, for that matter), yet there is still much controversy about whether these qualify as ``quantum'' computers and how accurately they presage eventual quantum computers that will deliver quantum supremacy \cite{Preskill1}. A term that is used to describe these prototypes is Noisy Intermediate-Scale Quantum (NISQ) hardware. Instead of manipulating single qubits, NISQ devices manipulate the flow of a large collection of qubits, cooled down to near absolute-zero, so that quantum and classical noise is suppressed and quantum effects like superposition manifest within the ensemble of qubits. This is in contrast to the theoretically more robust notion of quantum hardware which allows the manipulation of individual qubits to produce superposition and entanglement. Noise, in particular classical noise coming in from the environment, is again a fundamental challenge since programming a quantum processor necessarily requires interaction with a programmer residing in the environment. Some levels of noise can be tolerated if the hardware is built with error-detection and correction codes, an idea motivated by how noise induced errors are dealt with in classical hardware. However, implementing these codes in quantum hardware turns out to be an expensive endeavor \cite{Preskill}. Designing and realizing quantum hardware continues to be an active area of research both in academic and industrial settings as its successful implementation (with respect to noise suppression, or ideally, cancellation) has been shown to be the beginning of quantum supremacy over classical devices 

To this end, taking a mathematically formal approach, recall that the physics of qubits takes place in a complex projective space and then note that this space is also a compact Riemannian manifold. The physics of classical objects takes place in the Euclidean space, the stereotypical manifold of our every day experience. By John Nash's embedding theorem, we know that compact Riemannian manifolds can be embedded inside some high dimensional Euclidean space in a way that preserves length. We regard the embedded image of the compact manifold of qubits inside Euclidean space as representing exotic sectors in the Euclidean space that can be made to exhibit quantum properties by tracing the embedding back to its quantum origin. 

Nash embedding has the following appealing properties: as a submanifold of the Euclidean space, its image ``realizes'' qubits.
Being an isometry, that is, preserving length, it preserves relationships between qubits when they are realized. This means that the action of {\it any} quantum logic gate (unitary matrix) on the qubits can be faithfully implemented by some reversible logic gate (orthogonal matrix) acting on the image of the qubits, albeit in a Euclidean space with dimension possibly higher than the original space of qubits. This property is the main focus here. 

A noteworthy point here before proceeding to the details is that the classical environment is more properly psuedo-Riemannian than Euclidean, once relativistic considerations are taken into account. Psuedo-Riemannian (also known as semi-Riemannian) manifolds embed isometrically into the psuedo-Euclidean space (such as the Minkowski space) \cite{Clarke},\cite{Greene}, but the question of whether they isometrically embed into Riemannian manifolds appears to be unsettled. This is one explanation for why the reconciliation of quantum physics and relativity remains elusive. This issue will certainly come to the forefront when relativistic quantum circuits become ubiquitous, at which time new insights into reconciling the two forms of physics might become apparent.  

\section{Quantum state space and Nash embedding}

In a quantum computing context, the relevant compact Riemannian manifold is the two-dimensional complex projective space of a  single qubit, $\mathbb{C}P^{1}$. This set consists of equivalence classes of ``complex lines'', that is, all the non-zero vectors $v_1, v_2 \in \mathbb{C}^{2}$ declared equivalent if $v_1 =\lambda v_2$ for non-zero complex numbers $\lambda$. We can visualize this space by shrinking all the complex lines down to unit length so as to generate the unit sphere $\mathbb{S}^3$ in $\mathbb{R}^4 (=\mathbb{C}^2)$, and then identifying the antipodal points on $\mathbb{S}^3$. This means that 
\begin{equation}
\mathbb{C}P^1= \mathbb{S}^3/\mathbb{S}^1 = \mathbb{S}^2  \subset \mathbb{R}^3.
\end{equation} 
In quantum computing literature, $\mathbb{S}^2$ is utilized as the Bloch sphere, a {\it representation} of a qubit to assist with calculations.

Next, consider the Cartesian product $\mathbb{C}P^{1} \times \mathbb{C}P^{1}$ as the model for joint system of two qubits. This set can be given a four-dimensional Hilbert space structure using the direct-sum construction. However, the direct-sum is incompatible with projectivity. Instead, the tensor product can be utilized, which while not well-defined for projective spaces, is well-defined for Hilbert spaces. The model for the joint space of two qubits is therefore the projectified four-dimensional Hilbert space $\mathbb{C}^{2} \otimes \mathbb{C}^{2}=\mathbb{C}^{4}$, denoted as $\mathbb{C}P^{3}$. Using the Segre  embedding \cite{Bengtsson}, a copy of $\mathbb{C}P^{1} \times \mathbb{C}P^{1}$ can be found inside $\mathbb{C}P^3$ as a submanifold. Hence,  the image of the Segre embedding describes two qubit product states and the remaining $\mathbb{C}P^{3}$ describes two qubit entangled states. The joint complex projective space of $n>2$ qubits is produced by iterating this construction to get 
\begin{equation}\label{statespace}
\mathbb{C}P^{2^n-1} = \mathbb{S}^{(2^{n+1}-1)}/\mathbb{S}^1
\end{equation}
where $\mathbb{S}^{(2^{n+1}-1)}$ is the unit sphere in $\mathbb{R}^{2n}=\mathbb{C}^n$. A copy of the set 
$$
\mathbb{C}P^{1} \times \mathbb{C}P^{1} \dots \times \mathbb{C}P^{1}
$$
resides as a submanifold of $\mathbb{C}P^{2^n-1}$ and contains the product states of $n$ qubits. This submanifold is sometimes written as 
\begin{equation}\label{product}
\otimes_{i=1}^n (\mathbb{C}P^{1})_i
\end{equation}
which is an abuse of notation obviously but has the advantage of being a clear reference to the product. 

In \cite{nash}, Nobel Laureate John Nash established the following result:

\vspace{3mm}
\noindent {\bf Nash embedding theorem}: {\it For every compact Riemannian manifold $M$,  there exists an isometric embedding of $M$ into $\mathbb{R}^m$ for a suitably large $m$}. 
\vspace{3mm}

An embedding is a differentiable {\it homeomorphism}, that is, a bi-continuous one-to-one and onto function from the manifold onto a submanifold of $\mathbb{R}^m$. According to Gunther \cite{Gun}, 
\begin{equation}
m={\rm max}\left(\frac{k(k+5)}{2}, \frac{k(k+3)}{2}+5 \right)
\end{equation} 
where $k$ is the dimension of the $M$. Nash's result has been developed further over the years and a more up to date exposition can be found in \cite{Han}. In the next section, Nash embedding is applied to the $n$ qubit register $\mathbb{C}P^{2^n -1}$.

\subsection{Quantum logic gates as faithful reversible ones}\label{faith}
\begin{figure}[t] 
\begin{center}
 \begin{tikzcd}[column sep=huge, scale=1.2,transform shape]
\mathbb{C}P^{2^n -1} \arrow[hookrightarrow]{d}{e} \arrow{r}{Q} & \mathbb{C}P^{2^n -1}  \arrow[hookrightarrow]{d}{e} \\ 
S \arrow[densely dotted]{r}{R} & S
\end{tikzcd}
\caption{Quantum logic gate $Q$ transforming a quantum state to another. If $e$ is a Nash embedding into $S \subset \mathbb{R}^k$ for a suitable $k$, then the isometry of both means that there always exists an orthogonal transformation $R$ from $S$ to itself that realizes $Q$. }
\label{fig} 
\end{center}
\end{figure}

In \cite{Bennett}, Bennett showed that it is possible to make any logically irreversible circuit (or its corresponding Turing machine), logically reversible; in other words, the function computed by the circuit can be made invertible. This further implied physical or thermodynamic reversibility, meaning that in principle, digital computers can be built that dissipate an arbitrary small amount of heat. Bennett's (and related) works motivated further studies in reversible computing, leading Ingarden to formulate the (Shannon) theory of quantum information \cite{Ingarden} and leading Feynman to propose the construction of quantum computers \cite{Feynman} to simulate complicated quantum systems with simpler one's. As I show here, the ability of Nash embedding to realize a quantum circuit with a reversible one completes the proverbial circuit.  

To initialize the $n$ qubit register $\mathbb{C}P^{2^n -1}$, a unitary (and hence isometric) transformation is enacted on it as a quantum logic gate
\begin{equation}
Q: \mathbb{C}P^{2^n -1} \longrightarrow \mathbb{C}P^{2^n -1} 
\end{equation}
to configure its state to the desired one. 
Under a Nash embedding, the $n$ qubit register maps to a submanifold $S$ of a Euclidean space $\mathbb{R}^{m}$,
\begin{equation}
e:\mathbb{C}P^{2^n -1} \hookrightarrow  S,
\end{equation}
with $k=2^{n}$ and $m=2^{2n-1}+3\cdot 2^{n-1}+5$ for $n=1, 2$ and $m=2^{2n-1}+5\cdot 2^{n-1}$ otherwise. Because $e$ is an isometry, $Q$ can {\it always} be enacted with respect to the realized register $S$ via an orthogonal transformation
$$
R: S \longrightarrow S.
$$ 
That is, for $p \in \mathbb{C}P^{2^n -1}$
\begin{equation}\label{commute}
R(e(p)) =e(Q(p)) 
\end{equation}
as depicted in the commutative diagram of Figure \ref{fig} and from which it follows that
\begin{equation}\label{Q}
Q=e^{-1}(R(e(p))). 
\end{equation}

For the simplest instance of Nash embedding, consider the ``trivial'' one where $\mathbb{C}P^1  \hookrightarrow \mathbb{R}^{10}$, and a $2 \times 2$ unitary matrix $Q \in {\rm SU}(2)$ is realized by a $10 \times 10$ orthogonal matrix $R \in {\rm SO}(10)$. A two qubit register will Nash embed into $\mathbb{R}^{19}$! Even more dramatically, $\mathbb{C}P^7 \hookrightarrow \mathbb{R}^{52}$ and  $\mathbb{C}P^{15} \hookrightarrow \mathbb{R}^{168}$. Nash embedding clearly requires a large amount of Euclidean space by virtue of being an isometry; compare with the non-isometric Whitney embedding \cite{Whitney} for which $\mathbb{C}P^1\hookrightarrow \mathbb{R}^5$. 

For an $r \times r$ unitary with $2r^2$ real parameters and an $s \times s$ orthogonal matrix with $s^2$ real parameters, the orthogonal matrix will be less costly, with respect to number of real parameters that define it, as long as $s^2 < 2r^2$. But for the $n \geq 3$ qubit register, $r=2^n$ and $s=\left( 2^{2n-1}+5 \cdot 2^{n-1} \times 2^{2n-1}+5 \cdot 2^{n-1} \right)$, and for these values $s^2 > 2r^2$. Hence, for any computationally meaningful number of qubits, the reversible gate will always be costlier than the unitary gate. However, given that Nash embedding is the only robust way to realize quantum logic gates, this non-negotiable price is unavoidable.

%


\section{Noise and quantum hardware implementation}

Since programming the qubit register via $Q$ is equivalent under the Nash embedding to programming the image of the qubit register in the classical space via $R$, classical noise arising from the programming effort will now affect a real register. This real register may further be reduced to a binary register and classical error-detecting and correcting codes can be built into $R$. The well established theory and applications of classical error detecting and correcting codes will suffice for quantum computing. 
The main focus here however is the physical implementation of quantum hardware in $\mathbb{R}^3$. Reversible gates and circuits, when represented as orthogonal matrices, define hardware graphs in Euclidean space and can therefore be embedded into $\mathbb{R}^3$ using graph embedding techniques that have been utilized in the past for implementing high-dimensional hardware architectures such as VLSI.

More precisely, once Nash embedding has faithfully realized $Q$ as the orthogonal matrix $R$, this matrix can be represented as a weighted adjacency matrix for a graph \cite{Biggs} $G_R$ in $\mathbb{R}^m$. The graph $G_R$ represents the hardware architecture that implements $R$ in $\mathbb{R}^m$. Finally, $G_R$ can in turn be embedded inside $\mathbb{R}^3$ since {\it any} graph can be embedded in this space \cite{Cohen}. To further manage this embedding of the hardware graph $G_R$ into $\mathbb{R}^3$, techniques like {\it book embeddings} can be utilized where the graph in $\mathbb{R}^3$ is laid out as stack of sheets (planes) which connects together along a common back-bone of the ``book'' \cite{Chung}. The theoretical plan of action would be 
$$
Q \longrightarrow R \longrightarrow (G_R \subset \mathbb{R}^k) \longrightarrow \mathbb{R}^{3} \longrightarrow ({\rm Book \hspace{1mm}Graph} \subset {\rm stacked} \hspace{1mm}\mathbb{R}^{2})
$$
before fabrication of quantum hardware can begin. 

\section{Conclusion}

Nash's embedding result is non-algorithmic in nature. In the context of imaging and learning algorithms, efforts in trying to algorithmize Nash's embedding have been made, for instance, in \cite{McQueen}, \cite{Verma}, and \cite{Zhong}. However, similar efforts in algorithmic implementations that take into account features of quantum information data, such as quantum measurement, seem not to have been explored. While a direct, forward construction of a Nash embedding from $\mathbb{C}P^{2^n-1}$ to $\mathbb{R}^m$ requires topological and differential geometric study, it may also be possibe to use quantum state tomography and purification techniques to produce an inverse construction. Quantum state tomography and parameter estimation techniques attempt to recreate a quantum state from the date collected from repeated measurements (sometimes in different basis) of several copies of the state \cite{alt}, \cite{Paris}, \cite{Olivares}. Formally, this is the function
$$
t: \Delta_l \longrightarrow \Delta(\mathbb{C}P^{2^n-1})  \longrightarrow \mathbb{C}P^{2^n-1}
$$
where $t$ goes from the simplex of probability distributions in $\mathbb{R}^l$ to the space of density matrices, and then, after purification, to the state space of pure quantum states. For instance, $t$ can be the {\it linear inversion} function which uses a conditional probabilistic version of Born's rule to estimate a pure quantum state from measurement data. Tomographical methods may not immediately produce an inverse isometric embedding and will likely require mathematical tuning, but they serve as good first approximations.

\section{Conflict of Interest}

The author states that there is no conflict of interest.

\end{document}